\documentclass[aps,pre,twocolumn,superscriptaddress,showpacs]{revtex4}

\usepackage[dvips]{graphicx}
\usepackage{amsmath}

\begin{document}

\title{Heterogeneous-$k$-core versus Bootstrap Percolation on
  Complex Networks}

\author{G. J. Baxter}
\email[]{gjbaxter@ua.pt}
\affiliation{Departamento de F\'isica, I3N, Universidade de Aveiro,
Campus Universit\'ario de Santiago, 3810-193 Aveiro, Portugal}

\author{S. N. Dorogovtsev}
\author{A. V. Goltsev}
\affiliation{Departamento de F\'isica, I3N, Universidade de Aveiro,
Campus Universit\'ario de Santiago, 3810-193 Aveiro, Portugal}
\affiliation{A. F. Ioffe Physico-Technical Institute, 194021
  St. Petersburg, Russia}
\author{J. F. F. Mendes}
\affiliation{Departamento de F\'isica, I3N, Universidade de Aveiro,
Campus Universit\'ario de Santiago, 3810-193 Aveiro, Portugal}

\date{\today}

\begin{abstract}
We introduce the heterogeneous-$k$-core, which generalizes the
$k$-core, and contrast it with bootstrap percolation.
Vertices have a threshold $k_i$
which may be different at each vertex. If a vertex has less than $k_i$
neighbors it is pruned from the  network. The heterogeneous-$k$-core is the
sub-graph remaining after no further vertices can be pruned.
If the thresholds $k_i$ are $1$ with probability $f$ or $k \geq 3$
with probability $(1-f)$, the process forms one branch of
an activation-pruning process which demonstrates hysteresis. The other
branch is formed by ordinary bootstrap percolation.
We show that there are two types of transitions in this
heterogeneous-$k$-core
process: the giant heterogeneous-$k$-core may appear with a continuous
transition and there may be a second, discontinuous, hybrid
transition. 
We compare critical phenomena, critical clusters and
avalanches at the heterogeneous-$k$-core and bootstrap percolation
transitions.
We also show that network structure has a crucial effect on these
processes, with the giant heterogeneous-$k$-core appearing immediately
at a finite value for any $f > 0$ when the degree distribution tends
to a power law $P(q) \sim q^{-\gamma}$ with $\gamma < 3$.
\end{abstract}


\maketitle
       

Bootstrap percolation and the $k$-core are closely related concepts,
and in fact it is easy to confuse the two. 
Both belong to a new class of systems with hybrid phase
transitions, yet
 it can be clearly shown that the two processes do not
map onto each other. Here we elucidate the
relationship and differences between these two concepts by introducing
a generalization of the $k$-core, the heterogeneous-$k$-core.

The $k$-core  
is the maximal sub-graph whose vertices all have internal degree at
least $k$ \cite{Bollobas84}.
It has proved a useful tool giving
insight into the deep structure of complex networks  
\cite{Carmi2007, AlvarezHamelin06, AlvarezHamelin08, Dorogovtsev2006a, dgm2008}
, and has found applications
in diverse areas, from
rigidity \cite{Moukarzel2003} and jamming \cite{Schwarz2006}
transitions to real neural networks \cite{Chatterjee2008, Schwab} and
evolution \cite{Klimek2009} .  
The $k$-core has been extensively studied on tree-like networks,
starting with Bethe lattices \cite{Reich1978,Chalupa1979} and Random
graphs \cite{Pittel1996, Fernholz04, Molloy2005}, before finally being
extended to arbitrary degree distributions
\cite{Dorogovtsev2006a,Goltsev2006,Riordan2008, Corominas-Murtra2008}.
Hyperbolic lattices have also been considered \cite{Sausset2009}.
Other studies, mostly numerical, have considered the sizes of culling
avalanches \cite{Farrow2007,Shukla2008,Iwata2009}.
Results on non tree-like graphs have been largely numerical
\cite{Kogut1981,Parisi2008}, although some analytic results
incorporating clustering have recently been obtained
\cite{Gleeson2009, Gleeson2010}.
At the same time, bootstrap
percolation has emerged as a useful model for a variety of
applications such as neuronal activity \cite{Eckmann2007,Soriano2008,
  gdam09},
jamming and rigidity transitions and glassy dynamics
\cite{Sellitto2005,Toninelli2006}, and magnetic systems
\cite{Sabhapandit02}. In bootstrap percolation, a set of seed vertices
is initially activated, and other vertices become active if they have
$k$ active neighbors.
This process has been investigated on two and three
dimensional lattices (see \cite{Holroyd2003, Holroyd2006, Balogh2006,
  Cerf1999} and references therein).
Bootstrap percolation has been studied on the random
regular graph \cite{Balogh2007,Fontes2008}, on infinite trees
\cite{BaloghPeres06}, and most recently general complex networks
\cite{BDG1}.  Finite random graphs have also been studied
\cite{Whitney2009}. An interesting alternative formulation is the
 Watts model of opinions, in which the threshold is defined as a certain
 fraction of the neighbors rather than an absolute number  \cite{Watts2002}.
These processes may also be generalized so that the
thresholds may be different at each vertex \cite{BDG1,Gleeson2008}. 

Here we introduce a generalization of the $k$-core, the
heterogeneous-$k$-core.
In the heterogeneous-$k$-core,
each vertex $i$ in a network has a hidden
variable, its threshold value $k_i$. 
The heterogeneous-$k$-core is the
largest subgraph whose members have at least as many neighbors as
their threshold value $k_i$. This may include finite clusters as well
as any giant component.
If the $k_i$ are all equal we recover the
standard $k$-core. 
We define a simple representative example of the heterogeneous-$k$-core
(HKC) in which vertices have a threshold of either $1$ or $k \geq
3$, distributed randomly through the network with probabilities $f$
and $(1-f)$ respectively. This can be directly contrasted with
bootstrap percolation (BPC), in which vertices can be of two types: with
probability $f$ they are `seed' vertices which are always active, while
with probability $(1-f)$ vertices become active only if their number
of active neighbors reaches a threshold $k$. The difference between
these two processes arises because bootstrap percolation is an
activation process, beginning from a sparsely activated network, while
the heterogeneous-$k$-core is a pruning process \cite{Pittel1996, Fernholz04},
beginning from a
complete graph. It is thus possible to think of these two processes as
two branches of a hysteresis loop in an activation-pruning process.

We observe two transitions in the size of the giant
heterogeneous-$k$-core (giant-HKC): a continuous transition similar to
that found in ordinary
percolation, and a discontinuous, hybrid, transition, similar to that
found for the ordinary $k$-core. We find a complex
phase diagram for this giant-HKC with respect to the proportion of each
threshold and the amount of damage to the network, in which, depending
on the parameter region, either transition may occur first.
Two similar transitions are
observed in the phase diagram of the giant
component of active vertices in bootstrap
percolation (giant-BPC).

Finally, we show that network heterogeneity plays an important
role.  When the second moment of the degree distribution is
finite but the third moment diverges, the giant-HKC (or giant-BPC)
appears at a
finite threshold but not linearly, instead being a higher order
transition. When the second moment of the degree distribution diverges
-- as
in scale-free networks -- the thresholds may disappear completely, so
that the giant-HKC (or giant-BPC) appears discontinuously at a finite
value for any $f>0$ or $p>0$.

\section{The Heterogeneous-$k$-core and Bootstrap
  Percolation \label{comparison}}

\begin{center}
\begin{figure}[htb]
\includegraphics[width=0.45\textwidth]{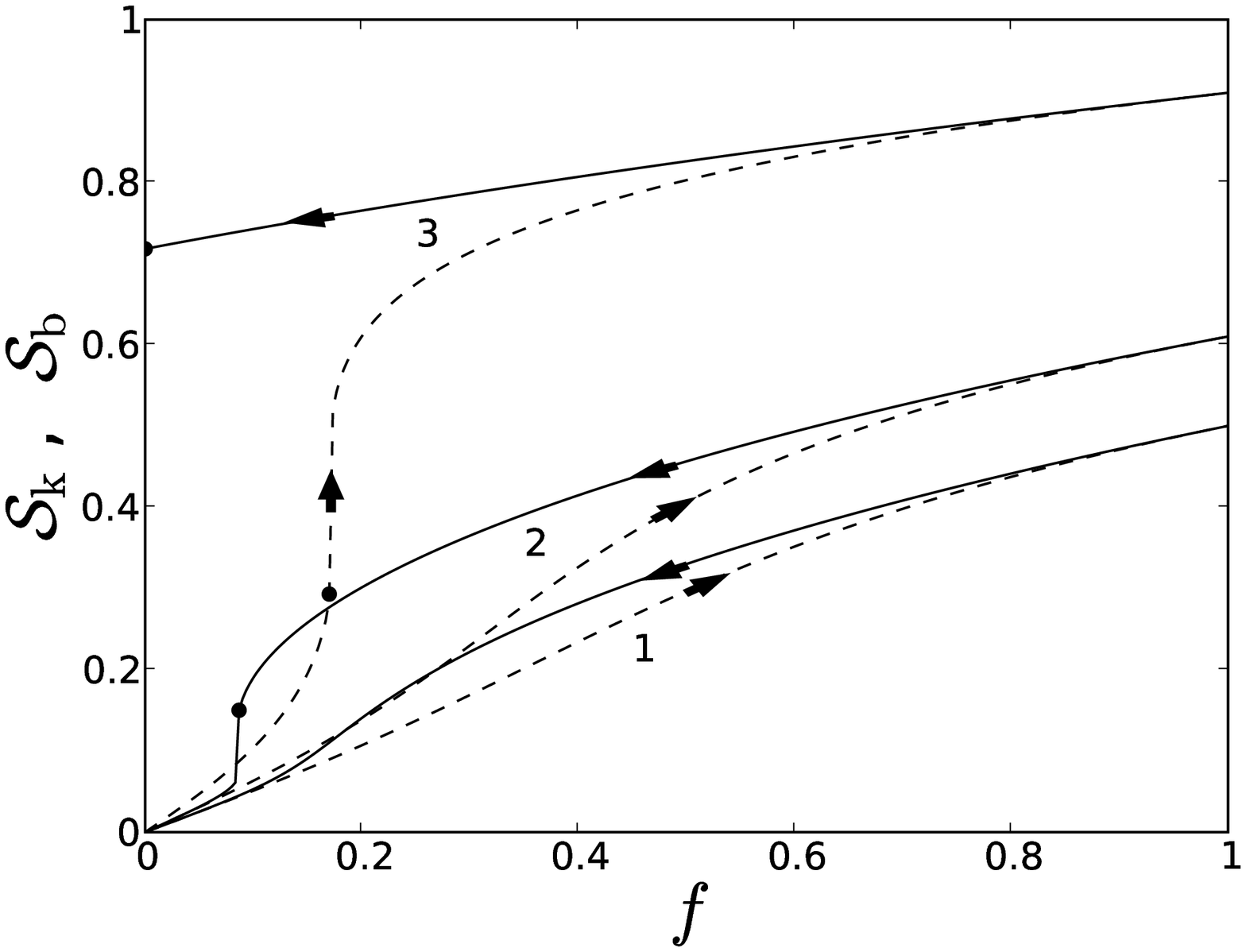}
\includegraphics[width=0.45\textwidth]{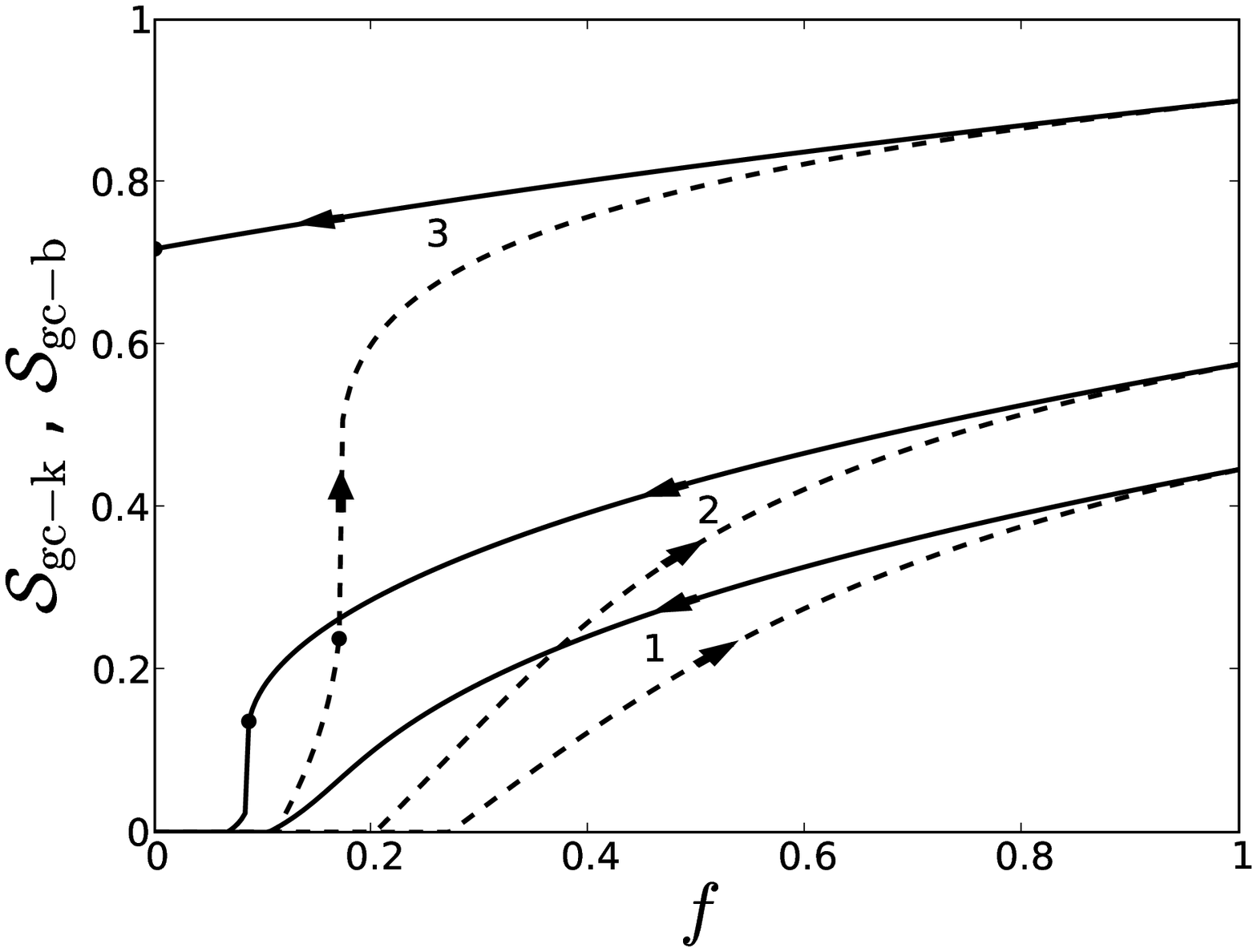}
\caption{Top: relative size $\mathcal{S}_{\text{k}}$ of the
  heterogeneous-$k$-core (solid curves), which is the subgraph
  including all vertices which  meet the threshold 
  requirements of Eq. (\ref{gencore_def}), and fraction
  $\mathcal{S}_{\text{b}}$ of active vertices in bootstrap percolation
  (dashed curves) as a function of $f$ for the
  same network -- an Erd\H{o}s-R\'{e}nyi graph of mean degree
$5$ -- with the same $k=3$, at three different values of $p$,
  corresponding to different regions of the phase diagrams
  Fig. \ref{phase_diag}.
1) $p = 0.5$, which  is between $p_{\text{c}}$ and $p_{\text{s}}$ for
both models. 2) $p = 0.61$, which is above $p_{\text{s-k}}$  but still
below  $p_{\text{s-b}}$. 3) $p=0.91$, which is above $p_{\text{f-k}}$
and $p_{\text{s-b}}$. Each numbered pair forms the two branches of a
hysteresis process, followed in the direction marked by the arrows.
Bottom: Size $\mathcal{S}_{\text{gc-k}}$ of the giant
heterogeneous-$k$-core (solid) and size
$\mathcal{S}_{\text{gc-b}}$ of 
the giant-BPC (dashed) as a function of $f$ for the same network
and the same values of $p$.
}\label{traces_both}\label{traces_genkcore}
\end{figure}
\end{center}

Consider an arbitrary, uncorrelated, sparse complex network, defined
by its degree distribution $P(q)$. In the infinite size limit, such
networks are locally tree-like, a property which enables the
analysis we will use.
 The network may be damaged to some extent by
the removal of vertices uniformly at random. The fraction of surviving
vertices is $p$.

In the heterogeneous-$k$-core, each vertex of a network is assigned a
variable $k_i \in \{0,1,2,...\}$. The $k_i$ values are assumed to be
uncorrelated, selected from a distribution $Q_k(r)$.
The heterogeneous-$k$-core is then the largest subgraph
of the network for which each vertex $i$ has at least $k_i$ neighbors
within the heterogeneous-$k$-core. 
To find the heterogeneous-$k$-core of a given network, we start with the
full network, and prune any vertices whose degree is less than its
value of $k_i$. 
As a result of this pruning, other vertices will lose
neighbors, and may thus drop below their threshold, so we repeat 
the pruning 
until a stationary state is reached. The remaining sub-graph is the
heterogeneous-$k$-core. If it occupies a non-vanishing fraction of the
original network in the limit that the size of the network goes to
infinity, we say it is a giant heterogeneous-$k$-core (giant-HKC). 

If all $k_i = 1$, then the HKC is simply the connected
component of the network, and the giant-HKC is the giant
connected component, exactly as in ordinary percolation. As is well
known \cite{AlbertBarabasi02,Dorogovtsev2002,dgm2008} this appears
with a continuous transition at the
critical point $p_c = \langle q \rangle/[\langle q^2\rangle  - \langle
  q\rangle]$, where $\langle q^n \rangle = \sum_i q^n P(q)$.
If $k=2$, we again have a continuous transition, similar to ordinary
percolation.
If all $k_i$ are equal to $k \geq 3$, then we have the ordinary
$k$-core. 
In this case the giant $k$-core appears with a discontinuous
hybrid transition 
\cite{Dorogovtsev2006a, Dorogovtsev2006b, Goltsev2006,Schwarz2006}.

Let us briefly discuss the nature of hybrid phase transitions.
These transitions form a specific new kind of phase transition
which combines a discontinuity like a first order phase transition
with a critical singularity like a continuous phase transition.
In thermodynamics, where the changes of a control
parameter (e.g., decreasing/increasing temperature) are assumed to be
infinitely slow, a first order transition has no hysteresis. In
reality, the changes always occur with a small but finite rate, and
the `heating' and `cooling' branches of a first order transition do
not coincide, which indicates the presence of a metastable state and
hysteresis. The width of the hysteresis increases with this rate until
some limit, which corresponds to the limiting metastable state.  The
resulting limiting curve for the order parameter has a square-root
singularity at the breakdown point. For example, if we heat a
ferromagnet with a first order phase transition sufficiently rapidly,
the curve `magnetization $M$ versus temperature $T$' has a singularity
$M(T)-M(T_b-0)\propto \sqrt{T_b-T}$ at the breakdown point $T_b$ and
discontinuity, that is $M(T_b-0)>M(T_b+0)=0$. After the breakdown, there
is no singularity. The susceptibility also shows a singularity
$\chi(T<T_b)\propto 1/\sqrt{T_b-T}$ and has no singularity above
$T_b$. In this sense, the hybrid (mixed) transition is a limiting
metastable state for a first order phase transition.  
The important property is that the transition is asymmetrical. There
are critical fluctuations and a divergent correlation length on only
one side of the critical point.
Continuous phase transitions demonstrate critical fluctuations and
divergent correlation length on both sides of the transition.
In first order phase transitions there are no critical fluctuations
and correlation length is finite everywhere, including at the critical
point.

In general we might expect a combination of continuous and hybrid
 transitions. To this end we first consider the simple case in which
 the $k_i$ are distributed between two values, controlled by a
 parameter $f$. Specifically:
\begin{equation}\label{gencore_def}
Q_k(r) = \begin{cases}
   f & \text{if } r = 1,\\
   1-f & \text{if } r = k,\\
0 & \text{otherwise},
  \end{cases}
\end{equation}
for some integer $k \geq 3$.
This parameterized HKC has as its two limits ordinary percolation
($f=1$) and the original $k$-core ($f=0$).

We now contrast this model with bootstrap percolation.
In bootstrap percolation, 
with probability $f$, a vertex is a `seed' and is initially
active, and remains active. The remaining vertices (a fraction
$1-f$) become active if their number of active neighbors
reaches or exceeds a threshold value $k$. Once activated, a
vertex remains active. The activation of vertices may mean that new
vertices now meet the threshold criterion, and hence become
active. This activation process continues iteratively until a
stationary state is reached.
The seed and activated vertices in bootstrap percolation are analogous
to the threshold $1$ and threshold $k$ groups in the heterogeneous-$k$-core.
We
might expect then that the subgraph formed by the active vertices in
the stationary state of bootstrap percolation might be related to the
heterogeneous-$k$-core. In fact, the two subgraphs are necessarily
different, as we will describe in detail in Sec. \ref{clusters}.
Nevertheless, the two processes have some similar or analogous
critical behaviors.
Note that we are using a lower threshold of $1$ for the generalized
$k$-core, meaning isolated vertices are not counted as part of the
HKC. We could also use a lower threshold of $0$, which would include
more vertices in the HKC, but would yield an identical giant-HKC. For
this reason, we can compare with bootstrap percolation, in which the
seed vertices have effectively a threshold of $0$. 

Because bootstrap percolation is an activation process,
 while the
heterogeneous-$k$-core is found by pruning, we can characterize them
as two branches of the same process, with the difference
between the curves shown in Fig.~\ref{traces_both} indicating
hysteresis. 
Consider beginning from a completely inactive network (that may be
damaged so that some fraction $p$ of vertices remain). As
we gradually increase $f$ from zero , under the bootstrap percolation
process, more and more vertices become active (always reaching equilibrium
before further increases of $f$) until at a certain threshold value,
$f_{c1-b}$ a giant active component appears. As we increase $f$
further, the size of the giant-BPC traces the dashed curves shown in
Fig.~\ref{traces_both}. See also \cite{BDG1}. The direction of this
process is indicated
by the arrows on these curves. Finally at $f=1$ all undamaged vertices
are active.
Now we reverse the process, beginning with a fully active network, and
gradually reducing $f$, de-activating (equivalent to pruning) vertices
that fall below their threshold $k_i$ under the heterogeneous-$k$-core
process. As $f$ decreases, the solid curves in
Fig.~\ref{traces_both} will be followed, in the direction indicated
by the arrows. Notice that 
the size of the giant-HKC for given values of $f$ and $p$
is always larger than the giant-BPC. 
The explanation for this difference will be explored in the following
Section.

\begin{center}
\begin{figure*}[htb]
\includegraphics[width=0.40\textwidth]{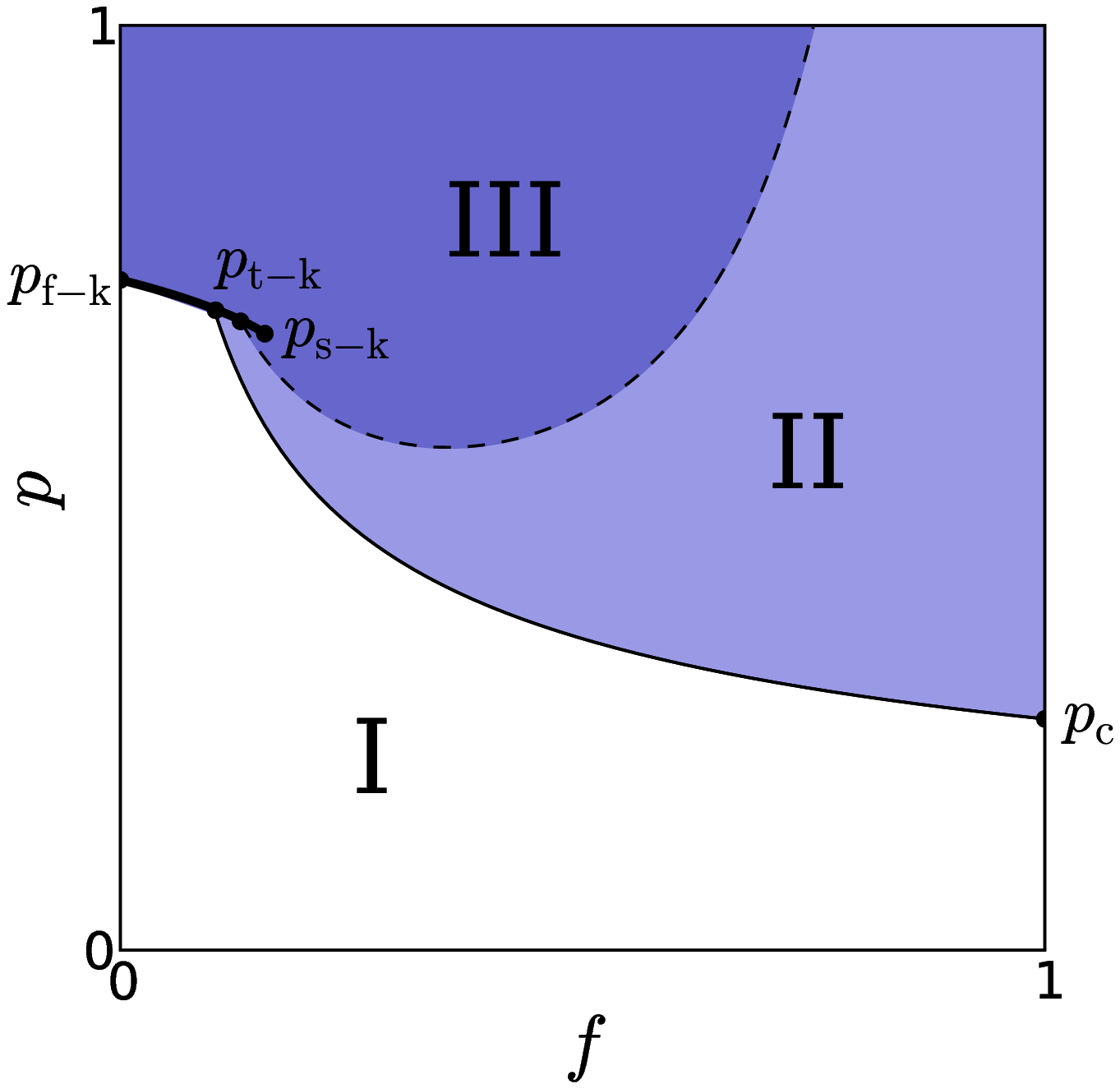}
\includegraphics[width=0.40\textwidth]{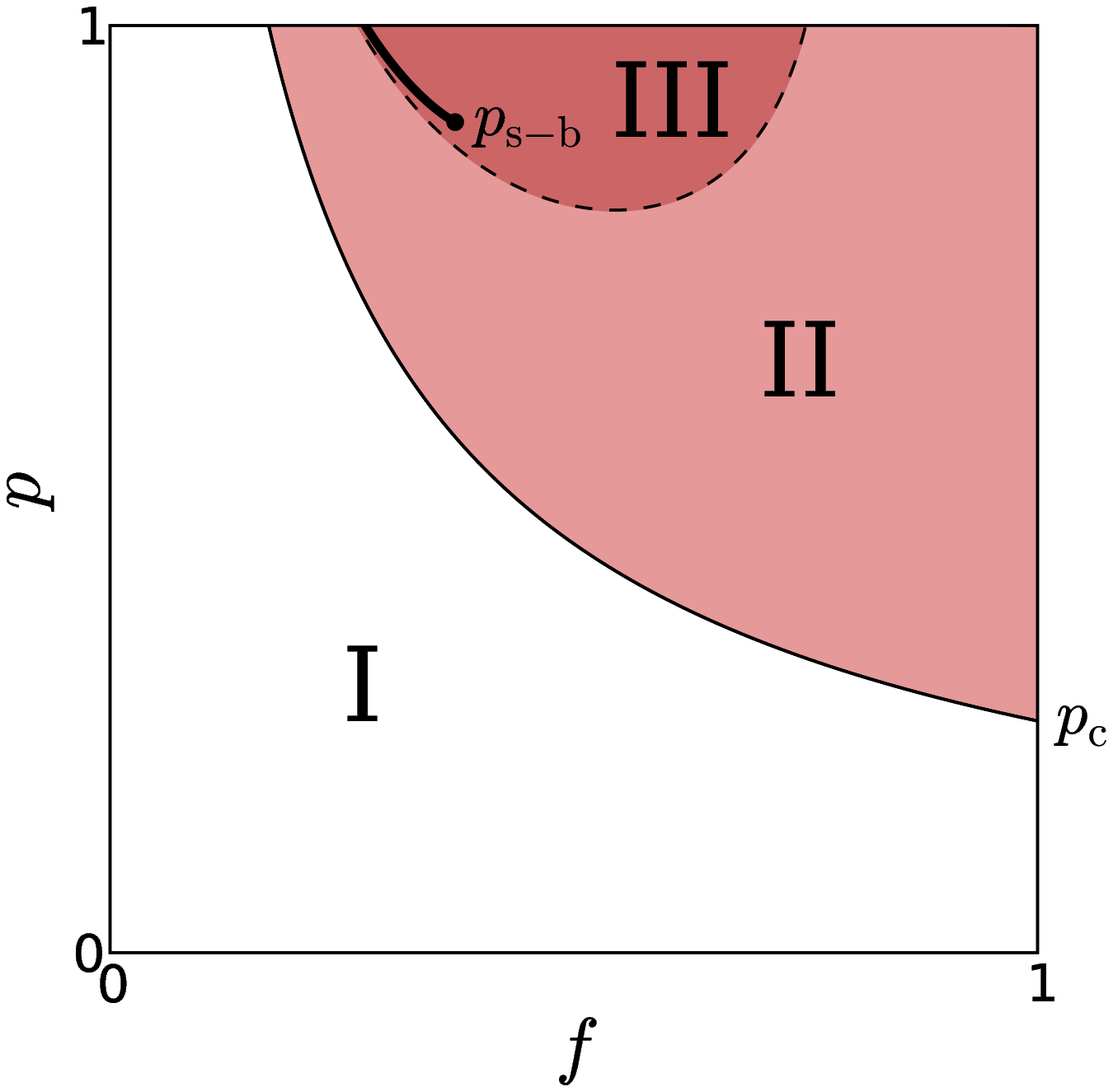}
\caption{Left: Phase diagrams for heterogeneous-$k$-core (left) in the
  $f$-$p$ plane. The giant-HKC is present in regions II and III,
  appearing continuously at the threshold marked by the thin
black curve.
The hybrid, discontinuous transition occurs at the
  points marked by the heavy black line, beginning from
  $p_{s-k}$.
Above $p_{t-k}$ the first appearance of the giant-HKC is
  with a discontinuous transition. Above $p_{f-k}$ the giant-HKC
  appears discontinuously for any $f>0$.
In region III the vertices with threshold $k$ in the HKC form a giant
connected component, which appears with a continuous
transition (dashed curve).
Right: Phase diagram for bootstrap percolation.
The giant-BPC is present in regions II and III. 
The continuous appearance of the giant-BPC is marked by the thin solid
curve, and the hybrid transition (beginning at $p_{s-b}$) by a heavy
solid curve. In region III the active vertices with threshold $k$ form
a giant connected component.
These diagrams are for a Bethe
  lattice with degree 5, and for $k=3$, but the diagram for any
  network with finite second moment of the degree
  distribution will be qualitatively the same.
Note that the locations of the continuous transitions from region I to
II and from II to III for bootstrap percolation with $k\to k-1$
coincide (up to the point where the discontinuous transition is
encountered) with those for the heterogeneous-$k$-core. The special
critical point $p_{s-b}$ with $k$ reduced by $1$ also coincides with
$p_{s-k}$ (see also Appendix \ref{AppSelfConsistency}).
}\label{phase_diag}
\end{figure*}
\end{center}

Let $\mathcal{S}_{\text{k}}$ be the fraction of vertices that are in the
heterogeneous-$k$-core. That is, the total of all components, whether
finite or infinite, that meet the threshold conditions.
This is equal to the probability that an arbitrarily
chosen vertex of the original network is in the
heterogeneous-$k$-core. Let $\mathcal{S}_{\text{gc-k}}$ be the relative
size
of the giant heterogeneous-$k$-core (that is, the subset of the
heterogeneous-$k$-core which forms a giant component) -- also the
probability that an arbitrarily chosen vertex is in the giant
heterogeneous-$k$-core. The fraction of vertices forming finite clusters
is therefore $\mathcal{S}_{\text{k}}-\mathcal{S}_{\text{gc-k}}$. Note
that in the standard $k$-core this is negligibly small.
Similarly, let $\mathcal{S}_{\text{b}}$ be the fraction of active
vertices in the bootstrap percolation model, and
$\mathcal{S}_{\text{gc-b}}$ be the size of the giant component of
active vertices.
We construct self-consistency equations for
$\mathcal{S}_{\text{k}}$ and $\mathcal{S}_{\text{gc-k}}$ 
in Appendix~\ref{AppSelfConsistency}.
In networks without heavy-tailed degree distributions, that is, whose
second moments do not diverge in the infinite size limit, we
find two different transitions for each process.

We first briefly describe the transitions observed in bootstrap
percolation, before comparing these with those found in the new
heterogeneous-$k$-core process.
For bootstrap percolation, above a certain value of $p$, the giant
active component (giant-BPC) may appear continuously from zero at a
finite value
of $f$, $f_{c1-b}$, and grow smoothly with $f$, see the dashed lines
$1$ and $2$ in the top panel of
Fig.~\ref{traces_both}. For larger $p$, after the
giant active component appears, there may also be a second
discontinuous hybrid phase transition, at $f_{c2-b}$, as seen in the
dashed line $3$ of
Fig.~\ref{traces_both}. There is a jump in the size of the giant
active component $\mathcal{S}_{\text{gc-b}}$ from the value at the
critical point (marked by a circle on dashed line $3$). When
approaching from
below, the difference of $\mathcal{S}_{\text{gc-b}}$ from the critical
value goes as the square root of the distance from the critical point:
\begin{equation}\label{scaling_bootstrap2}
\mathcal{S}_{\text{b}}(f) = \mathcal{S}_{\text{b}}(f_{\text{c2-b}}) -
a(f_{\text{c2-b}}-f)^{1/2}.
\end{equation}
See \cite{BDG1} for a complete description of the bootstrap
percolation results.

For the heterogeneous-$k$-core, we see a different but analogous pair of
transitions. Again, for a given $p$, the giant heterogeneous-$k$-core
(giant-HKC) appears continuously from zero above some critical value
of $f$, $f_{c1-k}$. There may also be a second transition, at
$f_{c2-k}$, where again we see a
discontinuity in the size of the giant-HKC. 
Now however, the square
root scaling occurs as the critical point is approached from above:
\begin{equation}\label{scaling_kcore2}
\mathcal{S}_{\text{k}}(f) = \mathcal{S}_{\text{k}}(f_{\text{c2-k}}) +
a(f - f_{\text{c2-k}})^{1/2}.
\end{equation}
See solid line $2$ in Fig.~\ref{traces_both}.
Another important difference is that, while
for bootstrap percolation the discontinuous transition is always above
the first appearance of the giant-BPC -- $f_{c2-b} > f_{c1-b}$, but
for the heterogeneous-$k$-core, $f_{c2-k}$ may be greater than $f_{c1-k}$
 -- so that
the first appearance of the giant-HKC is similar to that found in
ordinary percolation --  or less than $f_{c1-k}$, with the giant-HKC
appearing
discontinuously from zero, as in the ordinary $k$-core
\cite{Goltsev2006, Dorogovtsev2006b, Schwarz2006} (see
Fig.~\ref{traces_genkcore}).

The overall behavior with respect to the parameters $p$ and $f$ of
each model is summarized by the phase diagrams in 
Fig.~\ref{phase_diag}. 
The diagrams are qualitatively the same for
any degree distribution with finite second moment.
Considering first the heterogeneous-$k$-core, a giant-HKC is absent in
the region labelled I. For $p$ below the percolation threshold
$p_c$, the giant-HKC never appears for any $f$. Above $p_c$, the
giant-HKC appears with a continuous transition, growing linearly with
$f$ (or $p$ for that matter) close to the critical point. The
threshold is indicated by the thin black line in the Figure, which
divides regions I and II.
Compare line $1$ of Fig.~\ref{traces_genkcore}.
From $p_{s-k}$ a second, discontinuous, hybrid transition
appears. This is marked by the heavy black curve in
Fig.~\ref{phase_diag}. As already noted in
Eq.~(\ref{scaling_kcore2}) the size of the giant-HKC
grows as the square root of the distance above this second critical
point. At the special point $p_{s-k}$, however, the size of the
discontinuity reduces to zero, and the scaling near the critical point
is cube root:
\begin{equation}\label{scaling_kcore1}
\mathcal{S}_{\text{k}}(f) = \mathcal{S}_{\text{k}}(f_{\text{c2-k}}) +
a(f - f_{\text{c2-k}})^{1/3}.
\end{equation} 
At first the hybrid transition occurs after the continuous appearance
of the giant-HKC (see line $2$ of Fig.~\ref{traces_genkcore}), but at
$p_{t-k}$ the two transitions cross, and the
giant-HKC begins to appear immediately with a jump. Finally at
$p_{f-k}$, the giant-HKC begins to appear discontinuously immediately
from $f=0$ (line $3$ of Fig.~\ref{traces_genkcore}). Thus, the first
appearance of the giant-HKC has a classical
percolation like transition below $p_{t-k}$, while above $p_{t-k}$ the
appearance is similar to that found in the ordinary $k$-core.
Within the heterogeneous-$k$-core, the vertices with threshold $k$ may
form a giant component themselves. This occurs in the region labelled
III. This giant component appears with a continuous transition.

For bootstrap percolation, the giant active component again appears
only above the percolation threshold $p_c$. It also appears at first
with a linear, continuous transition (when the degree distribution has
finite third moment), but
at a larger value of $f$. The point of appearance is
marked by the thin curve in Fig.~\ref{phase_diag} which divides
regions I and II. Again, a
second transition appears for larger $p$, beginning at the special
critical point $p_{s-b}$. This transition is marked by the heavy solid
curve in the Figure. See also line $3$ of Fig.~\ref{traces_genkcore}.
The scaling near the hybrid transition is again
square root, but this time only when approaching the transition from
below. At the special critical point $p_{s-b}$, the scaling becomes
cube root:
\begin{equation}\label{scaling_bootstrap1}
\mathcal{S}_{\text{b}}(f) = \mathcal{S}_{\text{b}}(f_{\text{c2-b}}) -
a(f_{\text{c2-b}}-f)^{1/3}.
\end{equation}
Note the difference between Eqs. (\ref{scaling_bootstrap2}) and
(\ref{scaling_kcore2}) and between (\ref{scaling_bootstrap1}) and
(\ref{scaling_kcore1}). In bootstrap percolation, the hybrid
transition always occurs above the continuous one, and neither reaches $f=0$.
Again, the active threshold $k$ vertices form a giant component in
Region III.
Note also that the special critical point $p_{s-b} > p_{f-k}$, so that
for a given $p$ we may 
have a hybrid transition for the HKC or for BPC, but not for both.
It turns out that the special critical point $p_{s-b}$ for bootstrap
percolation whose value of $k$ is one less coincides with the special
critical point for the 
heterogeneous-$k$-core. 
Furthermore, the location of the continuous appearance of the giant
heterogeneous-$k$-core for a given $k$ also coincides with the
appearance of the giant component of bootstrap percolation for
$k-1$. The same is also true for the appearance of the giant
components of vertices with threshold $k$.
This is clear
from the equations given in Appendix~\ref{AppSelfConsistency}.
If the value of $k$ is increased, the locations of the hybrid
transitions move toward larger values of $p$, and there is a limiting
value of $k$ after which these transitions disappear altogether. For
both the heterogeneous-$k$-core and bootstrap percolation, this limit
is proportional to the mean degree. Note also that the continuous
transition also moves slightly with increasing $k$, and in the limit
$k\to\infty$, tends to the line $pf = p_c$ for both processes.

\section{Subcritical Clusters, Corona Clusters, Avalanches\label{clusters}}

It is clear from Figs. \ref{traces_both} and \ref{phase_diag} that
even though bootstrap percolation and the heterogeneous-$k$-core
described above have the same thresholds and proportions of each kind
of vertex, the equilibrium size of the respective giant components is
very different.
The difference results from the
top-down vs bottom-up ways in which they are constructed.
To find the heterogeneous-$k$-core, we begin with the full network, and
prune vertices which don't meet the criteria, until we reach
equilibrium. In contrast, bootstrap percolation begins with a largely
inactive network, and successively activates vertices until
equilibrium is reached. To see the effect of this difference, we now
describe an important concept: the subcritical clusters of
bootstrap percolation.

\begin{center}
\begin{figure}[htb]
\includegraphics[width=0.45\textwidth]{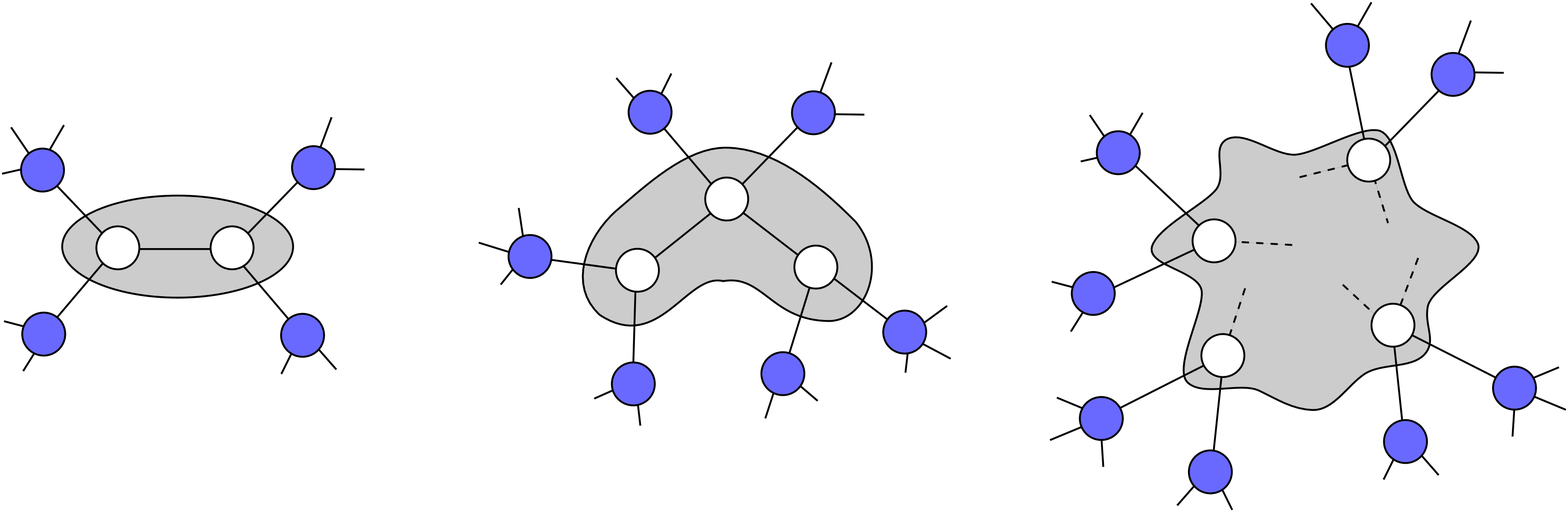}
\includegraphics[width=0.45\textwidth]{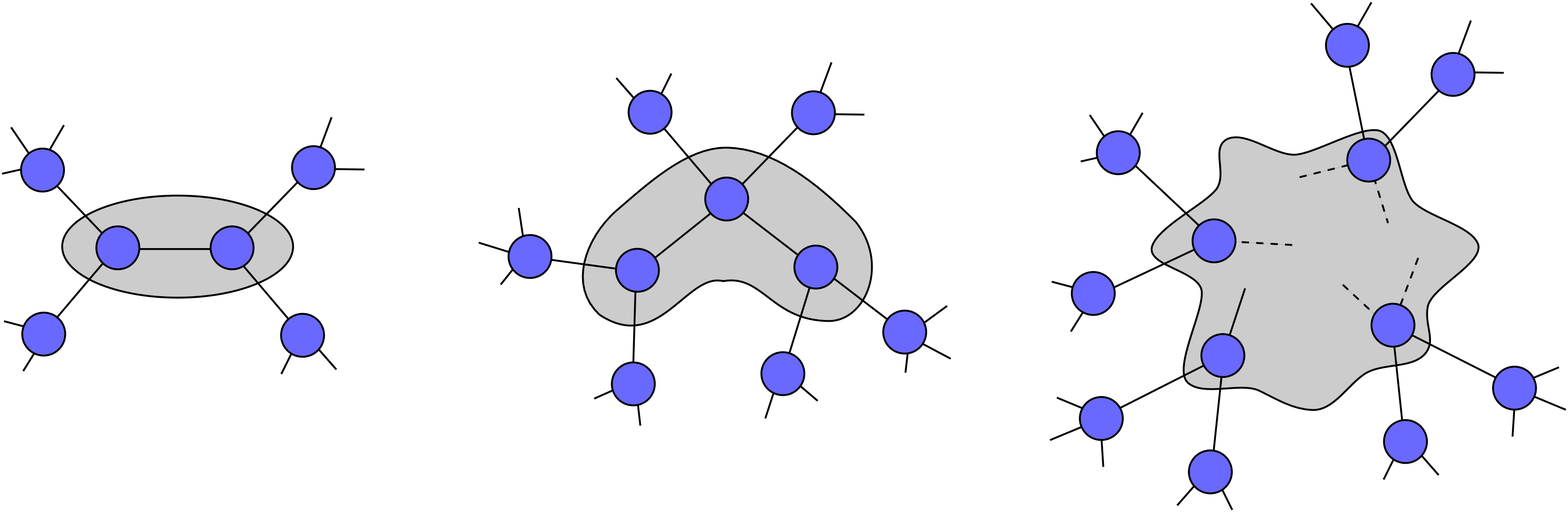}
\caption{Top row: Subcritical clusters of different sizes in bootstrap
  percolation. Left: Because they start in an inactive state, two
  connected vertices (shaded area) cannot become active if each has
  $k-1$ active neighbors. (In this example $k=3$.) 
The same follows for clusters of three (center) or more vertices (right).
If any member of a subcritical cluster gains another active neighbor, an
avalanche of activations encompasses the whole cluster.
Bottom row: Similar clusters would be included in the heterogeneous-$k$-core.
}\label{subcritical_clusters}
\end{figure}
\end{center}

\begin{center}
\begin{figure}[htb]
\includegraphics[width=0.45\textwidth]{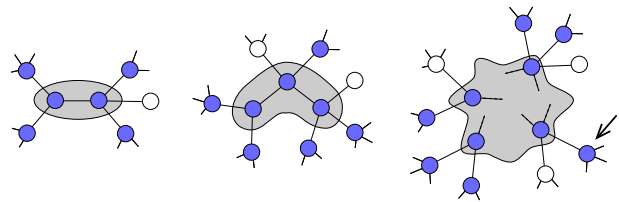}
\caption{Corona clusters of different sizes in
  heterogeneous-$k$-core. Left: Because they are included unless pruned,
  two connected vertices (whose threshold is $k$) form part of the
  heterogeneous-$k$-core if each has $k-1$ other neighbors in the core,
  as each is `assisted' by the other.
In general (right) a corona cluster consists of vertices (with
threshold $k$) that each
have exactly $k$ active neighbors, either inside or outside the
cluster. If one neighbor of any of the cluster vertices 
is removed from the core (for example, the one indicated by the
arrow), an avalanche is caused as the entire  cluster is
pruned.}\label{corona_clusters}
\end{figure}
\end{center}

A subcritical cluster in bootstrap percolation is a cluster of
activatable vertices (i.e. not seed vertices) who each have exactly
$k-1$ active neighbors external to the cluster. Under the rules of
bootstrap percolation, such clusters cannot become activated.
The vertices within the cluster block each other from becoming active
-- see Fig. \ref{subcritical_clusters}.
Now compare the situation for the heterogeneous-$k$-core.
Any cluster of threshold $k$ vertices which each have $k-1$ neighbors in
the core external to the cluster, is always included in the
heterogeneous-$k$-core. 
for the heterogeneous-$k$-core, vertices in clusters like those in
Fig. \ref{corona_clusters}
assist one another. 
Thus the exclusion of subcritical clusters from activation in
bootstrap percolation accounts for the difference in sizes of the
bootstrap percolation core and the heterogeneous-$k$-core. 

Subcritical clusters have another important property, that helps us to
understand the discontinuity at the second transition. 
A vertex in a subcritical cluster becomes active if it gains an extra
active neighbor (for example through an infinitesimal change in $p$
or $f$). This in turn allows each of its neighbors in the
cluster to activate. A domino-like effect ensues, leading to an
avalanche of activations -- one extra active neighbor of any vertex in the
subcritical cluster leads to the whole cluster becoming active. 
Thus
the rate of change of $\mathcal{S}_{\text{b}}$ is related to the sizes
of the subcritical clusters. Almost everywhere, if we choose a
subcritical vertex
at random, the mean size of the
subcritical cluster to which it belongs is
finite. However, exactly at the second
threshold, the mean size of the subcritical cluster to which a
randomly chosen vertex belongs diverges as we
approach from below. This was shown in \cite{BDG1}.
Thus,
approaching this point, an infinitesimal decrease in $f$ (or $p$) leads
to a finite fraction of the network becoming activated, hence a
discontinuity in $\mathcal{S}_{\text{b}}$ (and also in
$\mathcal{S}_{\text{gc-b}}$). The distribution of avalanche sizes near
the transition is determined by the size distribution $G(s)$ of subcritical
clusters, which at the critical point can be shown to follow $G(s)
\sim s^{-3/2}$. This can be shown using a generating function
approach, as demonstrated in \cite{BDG1}. A similar method can be
found in \cite{Goltsev2006, Dorogovtsev2006b,Callaway2000,Newman2001}.

We can then understand the hybrid transition in the
 heterogeneous-$k$-core, by considering the relevant clusters with
 similar properties,
 the corona clusters. Corona clusters are clusters of vertices with
 threshold $k$ that have exactly $k$ neighbors in the HKC. These clusters
are part of the heterogeneous-$k$-core, but if any member of a cluster
loses a neighbor, a domino-like effect leads to an avalanche as the
entire cluster is removed from the HKC.
 The corona clusters are finite
everywhere except at the discontinuous transition, where the mean size
of corona cluster to which a randomly chosen vertex belongs diverges
as we approach from above 
\cite{Goltsev2006,Dorogovtsev2006b,dgm2008}.
Thus, an
infinitesimal change in $f$ (or $p$) leads 
to a finite fraction of the network being removed from the
heterogeneous-$k$-core, hence a discontinuity in
$\mathcal{S}_{\text{k}}$ (and also in
$\mathcal{S}_{\text{gc-k}}$). The size distribution of corona
clusters, and hence avalanches at this transition also goes as a power
law with exponent $-3/2$.

\section{scale-free networks\label{scalefree}}

The results above, in Sec.~\ref{comparison} and in
Figs.~\ref{traces_both} and \ref{phase_diag},  are qualitatively the
same for networks with any degree
distribution which has finite second and third moments.
When only the second moment is finite, the phase diagram remains
qualitatively the same, but the critical behavior is changed. Instead
of a second order continuous transition, we have a transition of
higher order. When the second moment diverges, we have quite different
behavior.

To examine the behavior when the second and third moments diverge, we
consider 
scale-free networks, with degree distributions tending to the form
\begin{equation}
P(q) \approx q^{-\gamma}\;
\end{equation}
for large $q$. 
At present we consider only values of $\gamma > 2$.

To find the behavior near the critical points, we begin with a
self-consistency equation for $X$, which is the probability that
 an arbitrarily chosen edge leads to
an infinite $(k_i-1)$-ary tree (see Appendix
\ref{AppSelfConsistency}) 
The equation is given in the Appendix, as
Eq.~(\ref{X}).
 The probability $\mathcal{S}_{\text{gc-k}}$ can be
written in terms of $X$, and in fact both $X$ and
$\mathcal{S}_{\text{gc-k}}$ grow with the same exponent near the
appearance of the giant active component. 
We 
expand the right hand side of Eq.~(\ref{X}) near the appearance
of the giant-HKC (that is, near $X=0$). When $\gamma < 4$, the third
and possibly second moment of the degree distribution diverge. This
means that coefficients of integral powers of $X$ diverge, and we must
instead find leading non-integral powers of $X$. 

 When $\gamma > 4$, the second and third moments of the distribution
 are finite, so the behavior is the same as already described in
 Sec.~\ref{comparison}. Eq.~(\ref{X}) leads to
\begin{equation}\label{Xgamma_gt4}
X = c_1X + c_2X^{2} + \mbox{higher order terms}\;,
\end{equation}
which gives the critical behavior $X \propto (f -
f_{\text{c2-k}})^{\beta}$ with $\beta = 1$.
When $3 < \gamma \leq 4$, the linear term in the expansion of
Eq.~(\ref{X}) survives, but the second leading power is $\gamma-2$:
\begin{equation}\label{Xgamma4}
X = c_1X + c_2X^{\gamma-2} + \mbox{higher terms}\;,
\end{equation}
where the coefficients $c_1$ and $c_2$ depend on the degree
distribution, the parameters $p$ and $f$, and the (non-zero) value
of $Z$. The presence of the linear term means the giant-HKC appears at
a finite threshold, but because the second leading power is not $2$,
the giant-HKC grows not linearly but with exponent $\beta =
1/(\gamma-3)$. This means that the
phase diagram remains qualitatively the same as
Fig.~\ref{phase_diag}, however, the size of the giant-HKC grows as
$(f-f_{\text{c1}})^{\beta}$ 
with 
$\beta = 1/(\gamma -3)$. This is the same scaling as was found for
ordinary percolation \cite{Cohen2002}.

For values of $\gamma$ below $3$, the change in behavior is more
dramatic. 
When $2 < \gamma \leq 3$, the second moment of $P(q)$ also diverges,
meaning the leading order in the equation for $X$ is no longer linear
but $\gamma-2$:
\begin{equation}\label{Xgamma3}
X = d_1X^{\gamma-2}+ \mbox{higher terms}\; .
\end{equation}
From this equation it follows that there is no threshold for the
appearance of the giant-HKC (or giant-BPC). 
The
giant-HKC appears immediately and
discontinuously for
any $f>0$ (or $p>0$), and there is also no upper limit to the
threshold $k$.
This behavior is the same for bootstrap percolation, so the
(featureless) phase diagram is the same for both processes, even
though the sizes of the giant-HKC and giant-BPC are different.

\section{Discussion\label{discussion}}

We have introduced a new concept, the heterogeneous-$k$-core, an
extension of the well known $k$-core of complex networks.
A simple representative example of the heterogeneous-$k$-core (HKC) has
vertices with randomly assigned thresholds of either $1$ (with
probability $f$) or
$k\geq3$ (with probability $1-f$). This heterogeneous-$k$-core has a
complex phase diagram,
including two types of phase transition: a continuous transition at the
appearance of the giant heterogeneous-$k$-core, similar to the ordinary
percolation transition, and a second, discontinuous, hybrid phase
transition. This second transition is similar to that found for the
ordinary $k$-core, but it may occur after the first continuous
appearance of the
giant-HKC or before. The first transition occurs when the vertices of
both kinds reaching their threshold form a giant percolating
cluster.
The second transition occurs when the mean size of avalanches of
pruned vertices diverges.
This can be understood by considering 
corona clusters (clusters of vertices which exactly meet
the upper threshold). The 
mean size of the corona cluster to which an arbitrarily chosen vertex
belongs diverges as
we approach the second transition from above. The size of pruning
avalanches are determined by these corona clusters, and so the
transition is discontinuous.

We have contrasted the heterogeneous-$k$-core with bootstrap percolation,
in which there are also two kinds of vertices, but the core is defined
by an activation process, rather than a pruning process. The phase
diagram for bootstrap percolation therefore does not coincide with
that of the heterogeneous-$k$-core. Furthermore, the two processes can
be thought of as two branches of an activation-pruning process,
forming a hysteresis loop. The difference between the giant
heterogeneous-$k$-core and
the giant bootstrap percolation component results from the subcritical
clusters of bootstrap percolation. These clusters cannot be activated
as all members have $k-1$ active neighbors outside the
cluster. However, the equivalent clusters would be included in the
giant heterogeneous-$k$-core.

All of these results are strongly dependent on network structure. If
the third moment of the degree distribution is finite, we
obtain the results just described, with the giant-HKC growing linearly
above the continuous threshold. This is the case if the degree
distribution decays faster than a power-law $q^{-\gamma}$ with $\gamma
>4$ for large degree $q$. If instead the degree distribution tends to
a power-law with $3 < \gamma \leq 4$, so that the second moment is
finite while the third moment diverges, the phase diagram is
qualitatively the same, but the continuous transition is of higher
order.
If $2 < \gamma \leq 3$, the second moment of the degree distribution
diverges, and the situation is more extreme. The giant-HKC
appears immediately at a finite size for any $f>0$ or $p>0$, showing
that, in common with behavior found in other systems, such scale-free
networks are extremely resilient to damage.

\begin{acknowledgments} 

This work was partially supported by the following projects PTDC:
FIS/71551/2006, FIS/108476/2008, SAU-NEU/103904/2008, and
MAT/114515/2009, and also by the SOCIALNETS EU project. 

\end{acknowledgments} 

\appendix

\section{Self-Consistency Equations\label{AppSelfConsistency}}

Here we construct the self-consistency equations that the probabilities
$\mathcal{S}_{\text{k}}$, $\mathcal{S}_{\text{gc-k}}$,
$\mathcal{S}_{\text{b}}$, and $\mathcal{S}_{\text{gc-b}}$ must
obey. The (usually numerical) solution of these equations lead to the
phase diagrams and other results presented in
Section \ref{comparison}.
We have already given them for the case of bootstrap percolation
\cite{BDG1}.
\begin{table}[htb]
\includegraphics[width=0.24\textwidth]{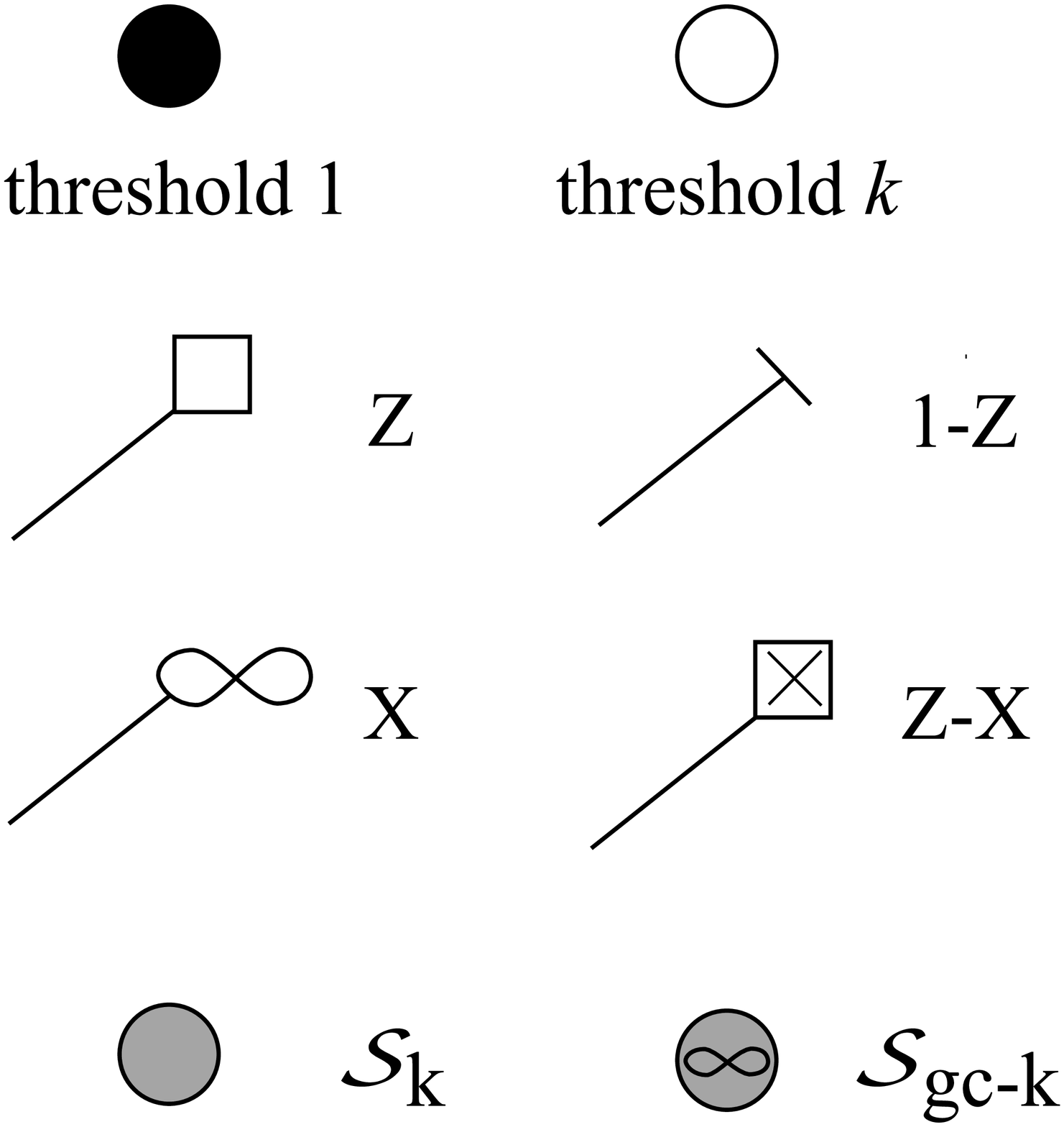}
\caption{Symbols used in graphical representations of self-consistency
  equations for the heterogeneous-$k$-core.\label{symbols_gencore}}
\end{table}

The probability $\mathcal{S}_{\text{k}}$ that an
arbitrarily chosen vertex belongs to the heterogeneous-$k$-core (HKC) is
the
sum of the probabilities that it has $k_i = 1$ and at least one
neighbor in the core, or $k_i = k$ and has at least $k$ neighbors in
the core. We can represent this diagrammatically as:
\begin{center}\label{feynman_Sk}
\includegraphics[width=0.29\textwidth]{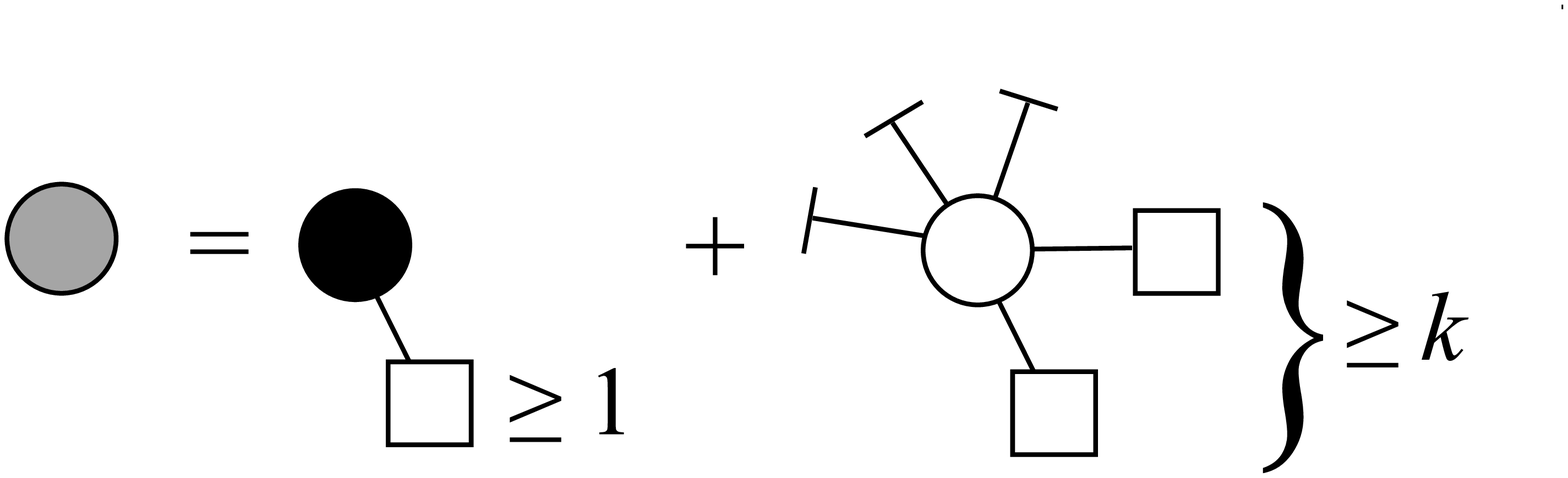}
\end{center}
We define $Z$ in terms of a `$(k_i-1)$-ary tree', a generalization of
the $(k-1)$-ary tree. A $(k_i-1)$-ary tree is a sub-tree in which, as
we traverse the tree, each vertex encountered has at least $k_i-1$
child edges (edges leading from the vertex, not including the one we
entered by). The variable $k_i$ can be different at each vertex. In
our example, $k_i$ is either $1$, in which case the vertex does not
need to have any children (though it may have them), and is only
required to be connected to the tree, or $k_i=k$, in which case it
must have at least $k-1$ children.

The probability $Z$ can then be very simply stated as the probability
that, on following an arbitrarily chosen edge in the network, we reach
a vertex that is a root of a $(k_i-1)$-ary tree. 
For the specific case considered in this paper, the
vertex encountered either has $k_i=1$, or it has $k-1$ children
leading to the roots of $(k_i-1)$-ary trees. The probability $Z$ is
represented by a square in the diagram.
A bar represents the probability $(1-Z)$,
a black circle represents a vertex with $k_i=1$, and a
white circle a vertex with $k_i=k$ -- see
Table~\ref{symbols_gencore}. These conditions can 
be written as binomial terms, and summing over all possible values of
the degree of $i$, this diagram can be written in
mathematical form as:
\begin{multline}
\mathcal{S}_{\text{k}} = pf\sum_{q=1}^{\infty}P(q)\sum_{l=1}^{q}
  \binom{q}{l}Z^l(1-Z)^{q-l}\\[5pt]
+ p(1-f)\sum_{q =  k}^{\infty} 
P(q)\sum_{l=k}^{q} \binom{q}{l}Z^l(1-Z)^{q-l}\;,
\label{Sk_full_app} 
\end{multline}
where the factor $p$ accounts for the probability that the vertex has
not been damaged.

To calculate $Z$, we construct a recursive (self consistency)
expression in a similar way, based on the definition given above. 
This is represented by the
diagram:
\begin{center}
\includegraphics[width=0.32\textwidth]{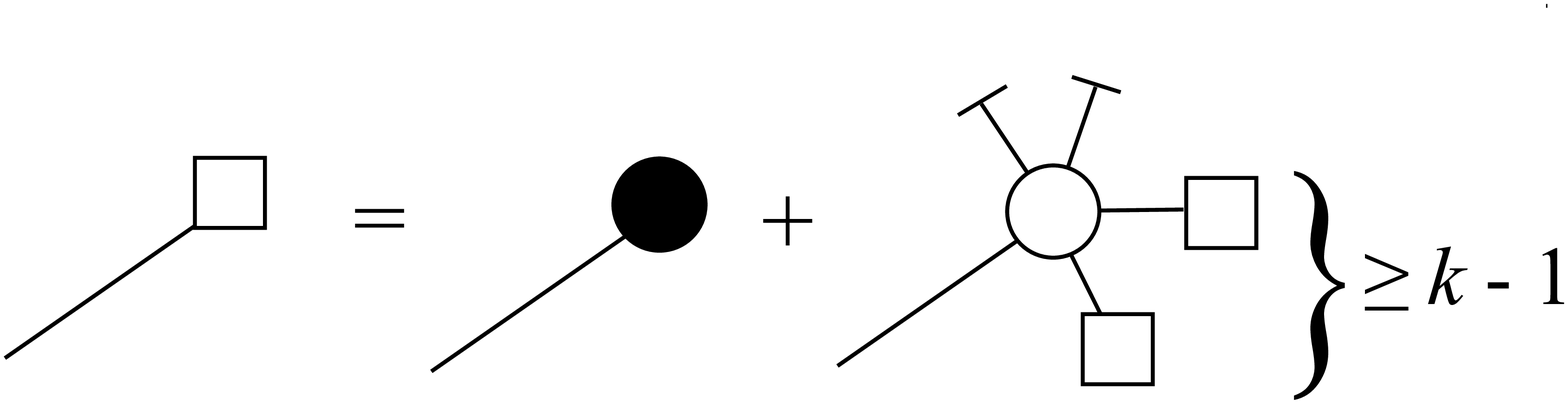}
\end{center}
which, in equation form is: 
\begin{align}
Z = & pf + \nonumber\\[5pt]
& p(1-f)\sum_{q\geq k}\frac{qP(q)}{\langle q\rangle} \sum_{l =
  k-1}^{q-1} \binom{q-1}{l} Z^l(1-Z)^{q-1-l}\nonumber\\[5pt]
\equiv & \Psi(Z,p,f)\;.
\label{Z}
\end{align}
We have used that $qP(q)/\langle q\rangle$ is the probability that the
vertex reached along an arbitrary edge has degree $q$.
Solving Eq.~(\ref{Z}) (usually numerically) for $Z$ and then
substituting into Eq.~(\ref{Sk_full_app}) allows the calculation of
$\mathcal{S}_{\text{k}}$.

We follow a similar procedure to calculate the size of the giant
heterogeneous-$k$-core (giant-HKC), which is equal to the probability
$\mathcal{S}_{\text{gc-k}}$ that an arbitrarily chosen vertex is a
member of a heterogeneous-$k$-core component of infinite size.
We denote by $X$ the probability that an arbitrarily chosen edge leads
to a vertex 
which is the root of an infinite $(k_i-1)$-ary tree. That is, the
definition is similar to $Z$, but with the extra condition that the
sub-tree reached must extend indefinitely.
 We represented $X$ by an infinity symbol
(Table~\ref{symbols_gencore}). The diagram for
$\mathcal{S}_{\text{gc-k}}$ is:
\begin{center}
\includegraphics[width=0.30\textwidth]{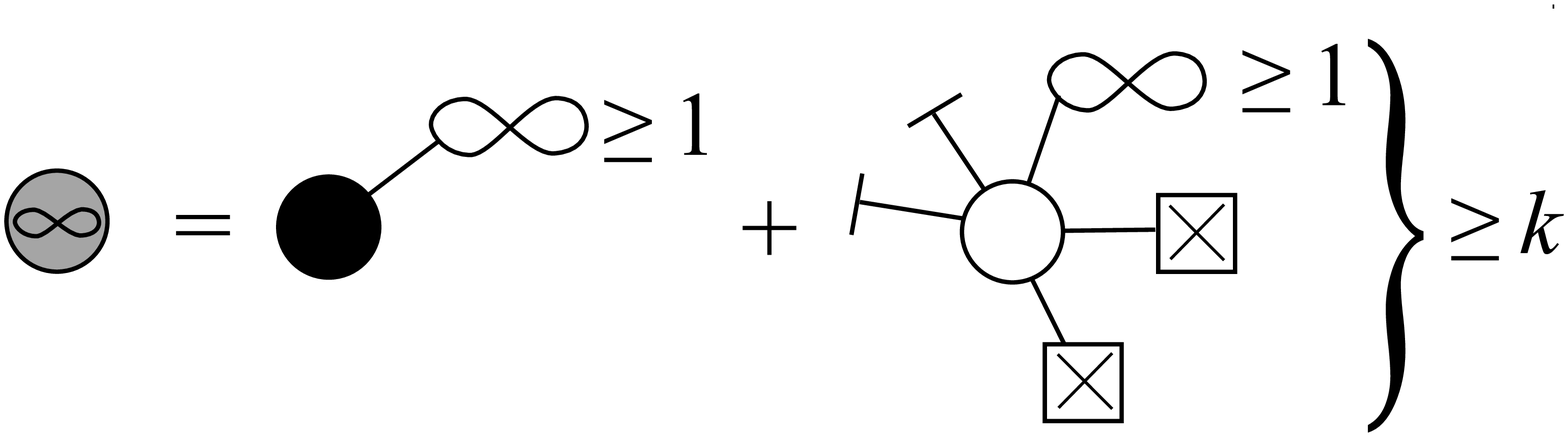}
\end{center}
which is equivalent to the equation:
\begin{align}
\mathcal{S}_{\text{gc-k}} =& pf \sum_{q=0}^{\infty} P(q)
\sum_{m=1}^q \binom{q}{m} X^m(1-X)^{q-m} \nonumber\\[5pt]
& +  p(1-f) \sum_{q=k}^{\infty}P(q) 
\sum_{l=k}^{q}\binom{q}{l}(1-Z)^{q-l}
\nonumber\\[5pt]
&\times \sum_{m=1}^{l}\binom{l}{m}
X^m(Z-X)^{l-m}\;.
\label{Sgc-k_full_app}
\end{align}

To find $X$, we construct a self-consistency equation from the diagram
\begin{center}
\includegraphics[width=0.41\textwidth]{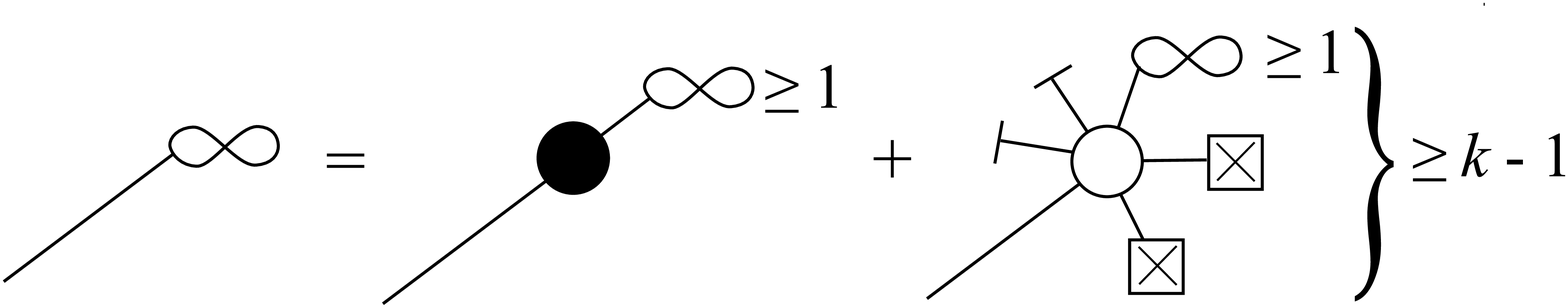}
\end{center}
leading to
 \begin{align}
X = pf & \sum_{q=0}^{\infty} \frac{q P(q)}{\langle q\rangle }
\sum_{m=1}^{q-1}\binom{q-1}{m} X^m(1-X)^{q-1-m}
\nonumber\\[5pt]
 +& p(1-f) \sum_{q=k}^{\infty}\frac{q P(q)}{\langle q\rangle }
\sum_{l=k-1}^{q-1}\binom{q-1}{l} (1-Z)^{q-1-l}\nonumber\\[5pt]
&{\times} \sum_{m=1}^l\binom{l}{m}X^m(Z-X)^{l-m}\;.
\label{X}
\end{align}
Solution of Eqs. (\ref{Z}) and (\ref{X}) then allows the calculation of
$\mathcal{S}_{\text{gc-k}}$ through Eq.~(\ref{Sgc-k_full_app}). Note that
when there are multiple solutions of $Z$, we choose the largest
solution as the `physical' one.

To find the appearance of the giant component for a given $p$, we find
leading terms for  $X \ll 1$, and solve for $f$, as described in
Sec.~\ref{scalefree}.
To calculate the location of the hybrid transition we note that
at this critical point a second solution to Eq.~(\ref{Z})
appears. This occurs when the function $\Psi(Z)$ just
touches the line $Z$, 
which must be at a local extremum of $\Psi/Z$:
\begin{equation}\label{Zjump}
\frac{d}{dZ}\left(\frac{\Psi}{Z} \right) = 0\;.
\end{equation}
Expanding Eq.~(\ref{Z}) about $Z_c$, the value of $Z$ at the critical
point (at the top of the jump), and using Eq. (\ref{Zjump}), we see
that $Z$ grows as the square-root of the distance from the critical
point. Using Eq.~(\ref{Sk_full_app}) we find
Eq.~(\ref{scaling_kcore2}).

Furthermore, at the special point $p_{s-k}$ where the second
transition disappears, by a similar argument, a further condition must
also be satisfied:
\begin{equation}\label{Zjump_last}
\frac{d^2}{dZ^2}\left(\frac{\Psi}{Z} \right) = 0.
\end{equation}
Thus, the critical point $p_{s-k}$ is determined by simultaneous
solution of Eqs. (\ref{Z}),(\ref{Zjump}), and (\ref{Zjump_last}).
This in turn leads to cube root scaling above the threshold, hence
Eq.~(\ref{scaling_kcore1}).

For bootstrap percolation, we can construct similar self-consistency
equations in order to calculate $\mathcal{S}_{\text{b}}$ and
$\mathcal{S}_{\text{gc-b}}$. Let $Y$ be the probability (counterpart
of $Z$)
 that on following an arbitrary edge, we encounter 
a vertex that is either a seed or has $k$ active children. 
As discussed in Section \ref{clusters}, activation in
bootstrap percolation must spread through the network, meaning that
the vertex needs $k$ active downstream neighbors in order to become
active (and thus provide an active neighbor to its upstream `parent').
Repeating the diagrammatic method described above, we arrive at the
equation:
\begin{align}
Y = & pf + \nonumber\\[5pt]
& p(1-f)\sum_{q\geq k+1}\frac{qP(q)}{\langle q\rangle} \sum_{l =
  k}^{q-1} \binom{q-1}{l} Y^l(1-Y)^{q-1-l}\nonumber\\[5pt]
\equiv & \Phi(Y,p,f)\;.\label{Y}
\end{align}
Note that Eq. (\ref{Y}) differs from (\ref{Z}) because the
number of children required is $k$ not $k-1$. This is equivalent to
excluding the subcritical clusters represented in Fig.
\ref{subcritical_clusters}. 
As an aside, consider the
probability $\mathcal{P}_{\text{sub}}$ that an arbitrary edge leads to a
vertex in a subcritical cluster. This is clearly simply the
probability that the vertex has exactly $k-1$ active neighbors, thus
\begin{equation}\label{p_subcritical}
\mathcal{P}_{\text{sub}} = p(1-f)\sum_{q \geq k} \frac{qP(q)}{\langle
  q\rangle} \binom{q-1}{k-1} Y^{k-1}(1-Y)^{q-k}\;.
\end{equation}
Comparing Eqs.~(\ref{Z}) and (\ref{Y}), we see that the right hand
side of Eq.~(\ref{p_subcritical}) contains precisely the terms that
are counted in Eq.~(\ref{Z}) but absent from (\ref{Y}).
Thus 
clusters of vertices all having
exactly $k-1$ active neighbors are excluded from the active component
in bootstrap percolation, while vertices having $k-1$ neighbors in the
heterogeneous-$k$-core are always included in the heterogeneous-$k$-core.
Of course because of the self-recursion, the value of $Z$ is
necessarily different from that of $Y$ by more than just the amount of
these terms.

Drawing diagrams similar to those given for the heterogeneous-$k$-core
allow the construction of further self-consistency equations for the
remaining quantities of interest. 
The probability that an arbitrarily chosen vertex is active,
$\mathcal{S}_{\text{b}}$ is then identical to Eq.~(\ref{Sk_full_app}) but
  with $Y$ replacing $Z$:
\begin{multline}
\mathcal{S}_{\text{b}} = pf\sum_{q=1}^{\infty}P(q)\sum_{l=1}^{q}
  \binom{q}{l}Y^l(1-Y)^{q-l}\\[5pt]
+ p(1-f)\sum_{q =  k}^{\infty} 
P(q)\sum_{l=k}^{q} \binom{q}{l} Y^l(1-Y)^{q-l}\;.
\label{Sb_full_app} 
\end{multline}
The equation for $\mathcal{S}_{\text{gc-b}}$ follows similarly. We
introduce the probability $W$ that, upon following an arbitrary edge,
we reach a vertex that is active and also has at least one edge
leading to an infinite active subtree.  Then $W$ obeys:
 \begin{align}
W = pf & \sum_{q=0}^{\infty} \frac{q P(q)}{\langle q\rangle }
\sum_{m=1}^{q-1}\binom{q-1}{m} W^m(1-W)^{q-1-m}
\nonumber\\[5pt]
 +& p(1-f) \sum_{q=k+1}^{\infty}\frac{q P(q)}{\langle q\rangle }
\sum_{l=k}^{q-1}\binom{q-1}{l}(1-Y)^{q-1-l} \nonumber\\[5pt]
&{\times} \sum_{m=1}^l\binom{l}{m}W^m(Y-W)^{l-m}\;,
\label{W}
\end{align}
which again, differs from Eq.~(\ref{X}) in that the limit is $k-1$ not
$k$. Then $\mathcal{S}_{\text{gc-b}}$ obeys:
\begin{align}
\mathcal{S}_{\text{gc-b}} =& pf \sum_{q=0}^{\infty} P(q)
\sum_{m=1}^q \binom{q}{m} W^m(1-W)^{q-m} \nonumber\\[5pt]
& +  p(1-f) \sum_{q=k}^{\infty}P(q) 
\sum_{l=k}^{q}\binom{q}{l}(1-Y)^{q-l}
\nonumber\\[5pt]
&\times \sum_{m=1}^{l}\binom{l}{m}
W^m(Y-W)^{l-m}\;,
\label{Sgc-b_full_app}
\end{align}
which is identical in form to Eq.~(\ref{Sgc-k_full_app}), though the
values of $Y$ and $W$ (for bootstrap percolation) will be different
from those of $Z$ and $X$ (for the heterogeneous-$k$-core). Note also
that, for bootstrap percolation the physical solution for $Y$ is
always the smallest of Eq.~(\ref{Y}).
The discontinuous transition occurs at the point where:
\begin{equation}\label{Yjump}
\frac{d}{dY}\left(\frac{\Phi}{Y} \right) = 0\;,
\end{equation}
and at the special point $p_{s-b}$ there is one more condition,
\begin{equation}\label{Yjump_last}
\frac{d^2}{dY^2}\left(\frac{\Phi}{Y} \right) = 0,
\end{equation}
where the function $\Phi(Y)$ is defined by Eq.~(\ref{Y}).

Finally we note that the appearance of the giant component of
threshold $k$ vertices in the heterogeneous-$k$-core process can be
found in a similar way to the appearance
of the giant-HKC. We define $R$ to be the probability that an
arbitrarily chosen edge leads to the root of an infinite subtree that
is a $(k_i-1)$-ary tree with all of the $k_i=1$ vertices removed. This
probability then obeys a self consistency equation similar to that for
$X$:
 \begin{align}
R = & p(1-f) \sum_{q=k}^{\infty}\frac{q P(q)}{\langle q\rangle }
\sum_{l=k-1}^{q-1}\binom{q-1}{l} (1-Z)^{q-1-l}\nonumber\\[5pt]
&{\times} \sum_{m=1}^l\binom{l}{m}R^m(Z-R)^{l-m}\;.
\label{R}
\end{align}
The appearance of the giant component of threshold $k$ vertices is
then found by expanding this equation to leading order with respect to
$R$ and solving for $f$. This leads to an equation which gives the
dashed line in Fig. ~\ref{phase_diag}. Recall that $Z(f,p)$ is determined by
Eq. ~(\ref{Z}).
A similar procedure yields the corresponding
transition in
bootstrap percolation.

\section{General Form of Equations\label{general}}

In this paper,
we have examined only a special case of the
heterogeneous-$k$-core, in which vertices have threshold either $1$ or
$k \geq 3$. For completeness, we now give the self-consistency
equations for arbitrary threshold distribution $Q(r)$.
The size $\mathcal{S}_{\text{k}}$ of the heterogeneous-$k$-core is 
\begin{align}
\mathcal{S}_{\text{k}} = p
\sum_{r\geq 1} Q(r)
\sum_{q =  r}^{\infty} P(q)
\left[\sum_{l=r}^{q} \binom{q}{l}Z^l(1-Z)^{q-l}\right],
\label{Sa_gen} 
\end{align}
where, as above, $Z$ is the probability of encountering a vertex $i$ 
which is the root of a $(k_i-1)$-ary tree:
\begin{align}
Z  =  p\sum_{r\geq1} Q(r)\sum_{q =  r}^{\infty}
\frac{(q)P(q)}{\langle q\rangle }\sum_{l=r-1}^{q-1}
\binom{q-1}{l}Z^l(1-Z)^{q-1-l}. \label{Zgen}
\end{align}
Similarly, the equation for the size of the giant active component is
\begin{align}
\mathcal{S}_{\text{gc-k}} 
= & p\sum_{r\geq1} Q(r)
\sum_{q=r}^{\infty}P(q)\sum_{l=r}^{q}\binom{q}{l}(1-Z)^{q-l}
\nonumber\\[5pt] 
&\times\sum_{m=1}^{l}\binom{l}{m}X^m(Z-X)^{l-m},
\label{Sgc_gen}
\end{align}
where $X$ obeys:
\begin{align}
X =& 
p\sum_{r\geq1} Q(r) \sum_{q=r}^{\infty}\frac{qP(q)}{\langle q\rangle }
\sum_{l=r-1}^{q-1} \binom{q-1}{l} (1-Z)^{q-1-l} \nonumber\\[5pt]
&{\times}
\sum_{m=1}^{l} \binom{l}{m}
X^m(Z-X)^{l-m}.
\label{Xgen}
\end{align}

We do not derive any results for this general case, but we can
speculate that a more complicated phase diagram would appear. If any
vertices have threshold less than $3$, i.e. $Q(1) + Q(2) > 0$, we
would find a continuous appearance of the giant-HKC. Thresholds of $3$
or more, on the other hand, contribute discontinuous transitions, 
and it may be that there are multiple such transitions.

\bibliography{network}

\end{document}